\documentclass[twocolumn,aps,prl,showpacs]{revtex4-1}

\usepackage{graphicx}
\usepackage{amsfonts}
\usepackage{amsmath}
\usepackage{amssymb}
\usepackage{xcolor}

\def\endproof{\vrule height6pt width6pt depth0pt}

\newcommand{\etal}{{\it{et al.}}}

\begin{document}


\title{Proposal of a two-qutrit contextuality test
free of the finite precision and compatibility loopholes}


\author{Ad\'an Cabello}
 \email{adan@us.es}
 \affiliation{Departamento de F\'{\i}sica
 Aplicada II, Universidad de Sevilla, E-41012 Sevilla, Spain}

\author{Marcelo Terra Cunha}
 \affiliation{Departamento de Matem\'atica, Universidade Federal de Minas Gerais,
 Caixa Postal 702, 30123-970, Belo Horizonte, MG, Brazil}


\date{\today}





\begin{abstract}
It has been argued that any test of quantum contextuality is
nullified by the fact that perfect orthogonality and perfect
compatibility cannot be achieved in finite precision
experiments. We introduce experimentally testable two-qutrit
violations of inequalities for noncontextual theories in which
compatibility is guaranteed by the fact that measurements are
performed on separated qutrits. The inequalities are inspired
by the basic building block of the Kochen-Specker proof of
quantum contextuality for a qutrit, despite the fact that their
proof is completely independent of it.
\end{abstract}


\pacs{03.65.Ud, 03.65.Ta, 03.67.Mn, 42.50.Xa}





\maketitle


{\em Introduction---}Contextuality is a fascinating property of
nature: The result of an experiment may depend on other
compatible experiments that may be performed. This is
surprising because the experiment gives the same result when
repeated after any number of compatible experiments. It is also
surprising because the probability of obtaining any particular
result does not depend on which compatible experiments might be
performed (probabilities are noncontextual). Nature's
contextuality was pointed out by the discovery that some
predictions of quantum mechanics (QM) cannot be reproduced by
any noncontextual theory, and that this conflict occurs for any
state of any system with three or more distinguishable states
\cite{KS67}. Quantum contextuality has far-reaching
consequences, such as the impossibility of describing nature by
classical theories with bounded speed of information
\cite{Bell64} or bounded density of memory \cite{KGPLC10}, or
the possibility of device-independent eternally secure
communications \cite{HHHHPB10}.

\begin{figure}
\centerline{\includegraphics[width=7.5cm]{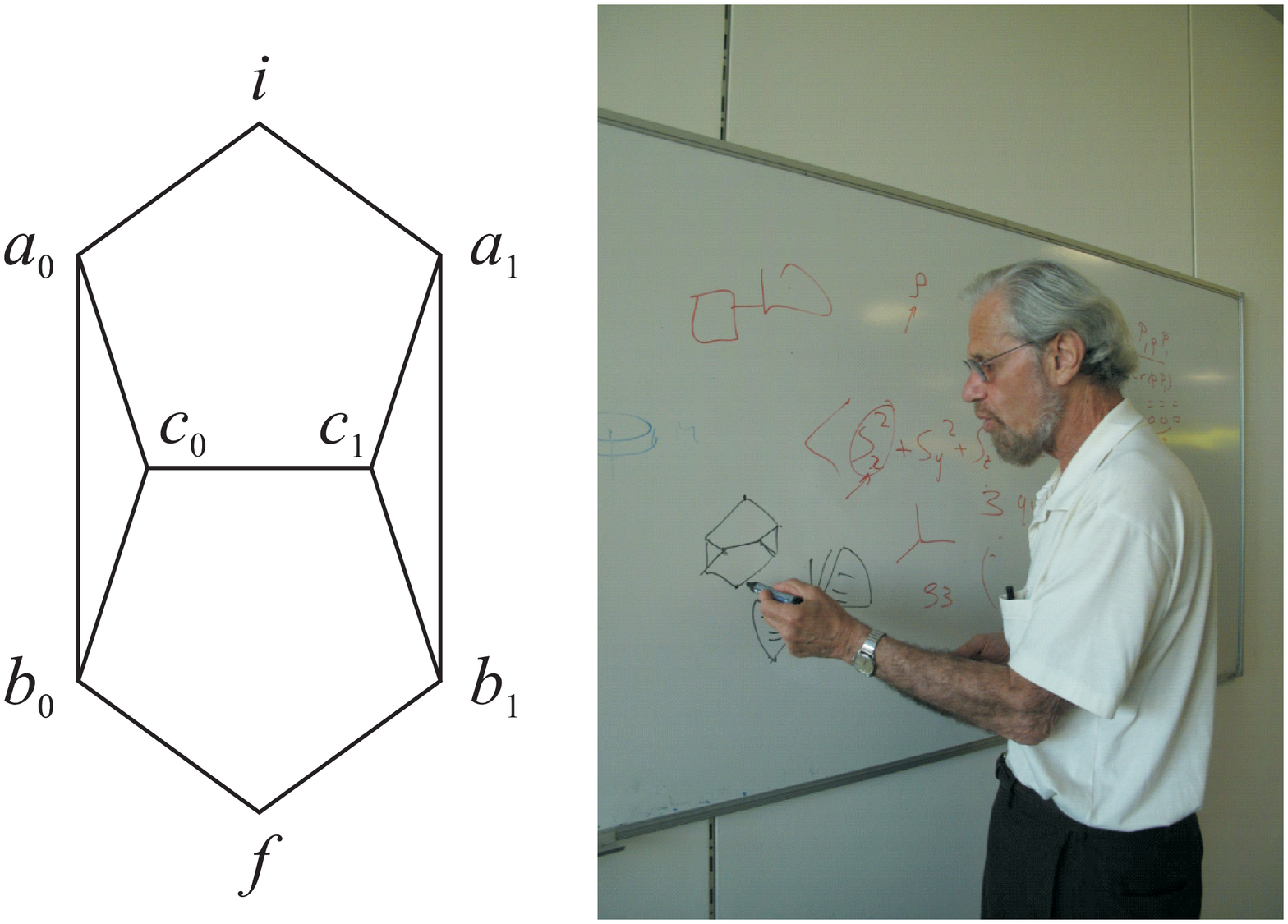}}
\caption{\label{KSblock} Building block of the Kochen-Specker
proof of quantum contextuality (left).
Vertices represent propositions; two of them are joined when the propositions
are compatible and both cannot be true.
Kochen drawing the block in Z\"urich in 2009 (right).}
\end{figure}

The original demonstration by Kochen and Specker (KS)
\cite{KS67} is usually considered a {\em{tour de force}}
because it is based on a $117$-vertex graph. However, this
graph has an $8$-vertex building block, which we refer to as
the basic diagram (Fig. \ref{KSblock}). In a KS graph, each
vertex is seen as a yes or no proposition and linked vertices
are compatible propositions where no more than one of them can
be assigned the yes answer (a property sometimes called
exclusiveness). The original $117$-vertex graph was built to
give a {state-independent} proof of the impossibility of
assigning noncontextual answers reproducing the predictions of
quantum mechanics for a three-level system (like a spin $1$).
Accordingly, a triangle represents a set of three compatible
propositions, one of them receiving the yes answer (given
quantum mechanically by a projective complete measurement, like
the $z$ component of a spin 1 particle). The basic building
block is also referred to as a yes-no diagram, in the sense
that if the answer yes is noncontextually assigned to vertex
$i$, any valid assignment must give to the vertex $f$ the value
no. This Letter was motivated by this basic diagram, as we will
show below.

In 1999, a debate on the physical impact and experimental
testability of the Kochen-Specker theorem started in the pages
of Physical Review Letters. Meyer \cite{Meyer99} and Kent
\cite{Kent99} pointed out that finite precision measurement
nullifies the physical content of the Kochen-Specker theorem.
More than 20 papers have been published since then supporting
(for instance, \cite{CK00, BK04}) or criticizing (for instance,
\cite{Mermin99, Cabello02}) this conclusion. More recently, a
series of experimental tests with ions \cite{KZGKGCBR09},
neutrons \cite{BKSSCRH09}, photons \cite{ARBC09, LHGSZLG09},
and nuclear magnetic resonance systems \cite{MRCL10} have been
questioned due to a variant of the objection of Meyer and Kent
called the compatibility loophole \cite{KZGKGCBR09,
GKCLKZGR10}.

In this Letter we present a proposal for a type of experiment
that (we hope) will definitely close the debate on the physical
relevance and experimental testability of the Kochen-Specker
theorem and stimulate a new generation of experiments.

Since it is impossible to have contextuality for a two-level
system, subjected to projective measurements, the three-state
quantum system, or qutrit, is considered the simplest example
of contextuality. We give here a state-dependent quantum
contextuality demonstration \cite{Stairs83, Clifton93} with a
twofold objective: to provide a clear example for the
nonspecialized reader and to fix notation to be used in the
two-qutrit case. Suppose the qutrit is initially in the state
\begin{equation}
|i\rangle=\frac{1}{\sqrt{3}}(|0\rangle+|1\rangle+|2\rangle),
\label{i}
\end{equation}
where $|0\rangle$, $|1\rangle$, and $|2\rangle$ are three
orthogonal states. Then the measurement represented by the
projector $|f\rangle\langle f|$, with
\begin{equation}
|f\rangle=\frac{1}{\sqrt{3}}(|0\rangle-|1\rangle+|2\rangle)
\label{f}
\end{equation}
gives the result $1$ (representing ``yes''), with (quantum mechanical) probability
\begin{equation}
|\langle f | i \rangle|^2 = \frac{1}{9}.
\label{oneovernine}
\end{equation}
This already establishes the {\it{yes-yes}} possibility and we
need only to prove the existence of the six other vertices
obeying all compatibility and exclusiveness properties of the
KS basic building block. For this, consider the observable
\begin{equation}
T_0 = a_0 |a_0 \rangle \langle a_0|+ b_0 |b_0 \rangle \langle b_0| + c_0 |c_0 \rangle \langle c_0|,
\label{H0}
\end{equation}
where $|a_0\rangle$, $|b_0\rangle$, and $|c_0\rangle$ are the
following orthogonal states:
\begin{subequations}
\label{h0}
\begin{align}
&|a_0 \rangle = \frac{1}{\sqrt{2}} (|1\rangle - |2\rangle), \\
&|b_0 \rangle = \frac{1}{\sqrt{2}} (|1\rangle + |2\rangle), \\
&|c_0 \rangle = |0\rangle.
\end{align}
\end{subequations}
Then, according to QM, the result $a_0$ can never be obtained
since $\langle a_0 | i \rangle =0$. In addition, in any
noncontextual theory in agreement with QM, the result of $T_0$
cannot be $b_0$ since $\langle f | b_0 \rangle =0$. Therefore,
in any noncontextual theory in agreement with QM, the initial
state $|i\rangle$ together with the positive probability
\eqref{oneovernine} imply $T_0=c_0$, whenever the yes answer
should be found for $|f\rangle$. Consider also
\begin{equation}
T_1 = a_1 |a_1 \rangle \langle a_1|+ b_1 |b_1 \rangle \langle b_1| + c_1 |c_1 \rangle \langle c_1|,
\label{H1}
\end{equation}
where
\begin{subequations}
\label{h1}
\begin{align}
&|a_1 \rangle = \frac{1}{\sqrt{2}} (|0\rangle - |1\rangle), \\
&|b_1 \rangle = \frac{1}{\sqrt{2}} (|0\rangle + |1\rangle), \\
&|c_1 \rangle = |2\rangle.
\end{align}
\end{subequations}
The same reasoning leads to the conclusion that, in any
noncontextual theory in agreement with QM, $T_1=c_1$ whenever
$f$ is to happen.

However, $T_0=c_0$ and $T_1=c_1$ cannot happen simultaneously
since $\langle c_1 | c_0 \rangle =0$. Therefore, there is no
noncontextual assignment of results to the six propositions
$|a_0\rangle \langle a_0|,\ldots,|c_1\rangle \langle c_1|$ in
agreement with the predictions of QM for a qutrit prepared in
the state $|i\rangle$ and postselected in the state
$|f\rangle$. Indeed, the orthogonality relations between these
eight states constitutes the building block of the
Kochen-Specker proof \cite{KS67, KS65}. These eight quantum
states obey all orthogonality conditions represented on the
building block diagram (see Fig.~\ref{KSblock}), but the
nonorthogonality of $|i\rangle$ and $|f\rangle$ allows for a
yes-yes answer, ruled out by noncontextual assignments.


What if perfect orthogonality cannot be experimentally
achieved? Then the previous argument vanishes. For instance,
the block of Fig.~\ref{KSblock} cannot be constructed if the
unit vectors are restricted to vectors with rational components
\cite{Meyer99} or to a set where each vector participates only
of one context \cite{CK00}, although in both cases these sets
are dense in the set of unit vectors and therefore are
undistinguishable from the set of all unit vectors by
experiments with finite precision \cite{Meyer99, Kent99, CK00,
BK04, Mermin99, Cabello02}. A related problem affecting some
recent experimental violations \cite{KZGKGCBR09, BKSSCRH09,
ARBC09} of noncontextual inequalities \cite{Cabello08, BBCP09}
involving sequential measurements on the same physical system
is the fact that these inequalities are based on the assumption
that the sequential measurements are perfectly compatible,
something which does not occur in actual experiments, so to
conclude contextuality extra assumptions are needed
\cite{KZGKGCBR09, GKCLKZGR10}.

Here we avoid these extra assumptions by measuring one of the
observables in one qutrit and the second observable on a
distant qutrit. Spatial separation provides a physical basis to
the assumption that both measurements are not only
approximately but perfectly compatible. The combination of
state-independent Kochen-Specker {proofs} and correlated
systems is a standard way to show contextuality and nonlocality
for composite systems \cite{Stairs83, Kochen, Cabello10,
AGACVMC11}. Here, instead, we do not require a
state-independent proof. Our central result will be to adapt
the previous one-qutrit example to find two-qutrit quantum
violations for noncontextual inequalities derived under the
same inspiration.

For this purpose we start by deriving an inequality valid for
any realistic theory (i.e., those assigning predefined answers
for all possible questions) based on the observables of the
block of Kochen and Specker and involving correlations between
measurements on two different parties $A$ and $B$.


{\em Lemma 1.---}Consider $D_j$ as dichotomic observables
{(with possible outcomes $1$ and $0$)}, $T_j$ as trichotomic
observables {(with possible outcomes $a_j$, $b_j$, and $c_j$)},
with $j=0,1$ and labels $A$ and $B$ corresponding to the
respective party. Then, for any realistic theory, the following
inequality holds:
\begin{subequations}
 \label{BHinequality}
\begin{equation}
{\cal K} \le 0,
\end{equation}
where
\begin{equation}
 \label{BHinequalityoperator}
\begin{split}
{\cal K}=&\sum_{j \neq k=0}^{1} P(D_j^A=1,D_k^B=1) -
 P(D_0^A=1,T_k^B=a_k)\\
 &-P(T_j^A=a_j,D_0^B=1)-P(D_1^A=1,T_k^B=b_k)\\
 &-P(T_j^A=b_j,D_1^B=1)-P(T_j^A=a_j,T_k^B=b_k)\\
 &-P(T_j^A=a_j,T_j^B=b_j)-P(T_j^A=b_j,T_j^B=a_j)\\
 &-P(T_0^A=a_0,T_0^B=c_0)-P(T_0^A=c_0,T_0^B=a_0)\\
 &-P(T_0^A=b_0,T_0^B=c_0)-P(T_0^A=c_0,T_0^B=b_0)\\
 &-P(T_j^A=c_j,T_k^B=c_k),
\end{split}
\end{equation}
\end{subequations}
where $P(D_0^A=1,T_0^B=a_0)$ denotes the joint probability of
obtaining the results $1$ and $a_0$ for $D_0^A$ and $T_0^B$,
respectively.


{\em Proof.---}Each probability in \eqref{BHinequality} can be
written as a sum of a set of probabilities for complete
hidden-variable states. In our scenario there are four
dichotomic observables, $D_0^A$, $D_1^A$, $D_0^B$, and $D_1^B$,
and four trichotomic observables, $T_0^A$, $T_1^A$, $T_0^B$,
and $T_1^B$. The proof is a bookkeeping exercise showing that
each term that adds to the positive term $P(D_j^A=1,D_k^B=1)$
also appears (with the right multiplicity) at some of the
negative terms, implying ${\cal K} \le 0$. A detailed
calculation is given in the supplemental material
\ref{suppmat}. \hfill\endproof

Notice that, despite the fact that the inequality
\eqref{BHinequality} is inspired by the KS diagram applied to a
pair of three-level systems in a singlet state, as will be made
clear by Eqs. \eqref{QMOperators}, it is independent of it in
the sense that the proof is not based on a diagram, and even a
computer programmed to search for realistic inequalities could
have found it.

It is also important to stress that inequality
\eqref{BHinequality} also holds for any local theory. This
means that, when the choice of measurement in one of the
particles is spacelike separated from the result of the
measurement on the other particle, then \eqref{BHinequality} is
a KS inspired Bell inequality.

In translating the KS basic diagram into four realistic
questions, exclusiveness of each triangle was used, but the
noncontextual assumption deserves more. For each part, we also
have the following logical implications:
{\begin{subequations} \label{compat}
\begin{align}
 &T_j=a_j \Rightarrow D_0=0, \\
 &T_j=b_j \Rightarrow D_1=0, \\
 &T_j=c_j \Rightarrow T_k \neq c_k\;(j\neq k).
\end{align}
\end{subequations}}
A noncontextual assignment is given by answers to the four
questions following the conditions \eqref{compat}. Only $14$
out of the $36$ realistic states are noncontextual. With fewer
allowed states, more restrictions can be obtained. One can now
prove the following.

{\em Lemma 2.---}Consider $D_j$ and $T_j$ as in Lemma 1.
Moreover, consider that these questions obey, for each party,
conditions \eqref{compat}. The following inequality holds:
\begin{align}
 \label{SingIneq1}
P_{\text{nc}}(D_0^A=1, D_1^B=1) - P_{\text{nc}}(D_0^A=1, T_0^B=a_0)& \\ - P_{\text{nc}}(D_0^A=1, T_1^B=a_1) & \leq 0.\nonumber
\end{align}
Where $P_{\text{nc}}$ is used to remind that only noncontextual
hidden-variable states are taken into account.

{\em Proof.---}Again, it is simple bookkeeping showing that all
terms which contribute to the positive term also appear at (at
least one) negative ones. Details and other inequalities are
provided in the supplemental material \cite{suppmat}.
\hfill\endproof

Inequalities \eqref{SingIneq1} are tight, with respect to
noncontextual states, and can be violated by realistic states
which do not obey some of the relations \eqref{compat}. In this
sense, even with an experiment prepared with spacelike
separations between $A$ and $B$, they should not be considered
as Bell inequalities. Their violation rules out a different set
of hidden-variable models: noncontextual ones. From the
technical point of view, that is a great advantage, since an
experimentalist need not bother with guaranteeing spatial-like
separations.

With the same kind of proof one can deduce the following.

{\em Lemma 3.---}With the same conditions of Lemma 2,
\begin{equation}
\label{SingIneq2}
\begin{split}
P_{\text{nc}}(D_0^A=1, D_1^B=1) &-\sum_j P_{\text{nc}}(T_j^A=b_j, T_j^B=a_j) \\
& - \sum _{j\neq k} P_{\text{nc}}(T_j^A=c_j, T_k^B=c_k) \leq 0.
\end{split}
\end{equation}
And also other similar expressions. Again, \eqref{SingIneq2} is
tight with respect to noncontextual states, and can be violated
with realistic contextual ones.


{\em Quantum violation.---}The same quantum system will violate
all inequalities shown here. It is just a two particle
generalization of the previous described state-dependent proof
of impossibility of noncontextual models for quantum mechanics.
Consider a two-qutrit system prepared in the state
\begin{equation}
|\psi\rangle = \frac{1}{\sqrt{3}} \left(|02\rangle - |11\rangle + |20\rangle\right),
\label{singlete}
\end{equation}
and the observables
\begin{subequations}
\label{QMOperators}
\begin{align}
&D_0^A=D_1^B = |i\rangle \langle i|,\\
&D_1^A=D_0^B = |f\rangle \langle f|, \\
&T_0^A=a_0 |a_0\rangle \langle a_0| + b_0 |b_0\rangle \langle b_0| + c_0 |c_0\rangle \langle c_0|,\\
&T_0^B=a_0 |b_1\rangle \langle b_1| + b_0 |a_1\rangle \langle a_1| + c_0 |c_1\rangle \langle c_1|,\\
&T_1^A=a_1 |a_1\rangle \langle a_1| + b_1 |b_1\rangle \langle b_1| + c_1 |c_1\rangle \langle c_1|,\\
&T_1^B=a_1 |b_0\rangle \langle b_0| + b_1 |a_0\rangle \langle a_0| + c_1 |c_0\rangle \langle c_0|,
\end{align}
\end{subequations}
with $|i\rangle$, $|f\rangle$, $|a_j\rangle$, $|b_j\rangle$,
and $|c_j\rangle$ as defined in Eqs \eqref{i}, \eqref{f},
\eqref{h0}, and \eqref{h1}. Then the predictions of QM are
\begin{equation}
 \label{QMProbs}
P(D_j^A=1,D_k^B=1)=\frac{1}{27},
\end{equation}
where $j,k \in \{0,1\}$ and $j \neq k$, and the remaining
probabilities in \eqref{BHinequalityoperator} are null.
Therefore, according to QM,
\begin{equation}
{\cal K}_{\rm QM} = \frac{2}{27} \approx 0.074,
\end{equation}
violating inequality \eqref{BHinequality}, and inequalities
\eqref{SingIneq1} and \eqref{SingIneq2} read
\begin{equation}
\frac1{27} \leq 0.
\end{equation}


{\em Discussion.---}Inequality \eqref{BHinequality} was deduced
using a translation of the basic KS diagram into a set of four
propositions. After this translation is done, its validity
rests only on hidden-variable hypothesis of a preexisting
reality which is reveled by measurements. Inequalities
\eqref{SingIneq1} and \eqref{SingIneq2} are based on these four
questions and on the constraints \eqref{compat} given by
noncontextual hypothesis. Because the lemmas were written and
proved, there is no necessity of going back to the diagrams,
despite the fact that their essence is captured by those
relations.

In all these inequalities, each term involves pairs of
compatible questions, asked for different parts of the composed
system. Moreover, being a quantitative statement, the argument
is not sensible to finite precision problems, given that
precision is enough to make violations meaningful. In this
sense those are proposals for quantum violations of
noncontextuality free from compatibility and finite precision
loopholes.


Testing the violation of inequality \eqref{BHinequality} is
very demanding experimentally. It requires testing $22$ joint
probabilities and, assuming that one obtains
$P(D_j^A=1,D_k^B=1)\approx \frac{1}{27}$ and $\epsilon$ for the
other $20$ probabilities, one must have $\epsilon <
\frac{1}{270} \approx 0.0037$ in order to observe a violation
of inequality \eqref{BHinequality}. The importance of this
result is that it explicitly shows that there is no fundamental
obstacle observing Kochen-Specker contextuality on (pairs of)
qutrits. No additional assumptions are needed to deal with the
fact that measurements have a finite precision and
compatibility between sequential measurements is not perfect.

Inequalities \eqref{SingIneq1} are less demanding experimentally. Indeed,
they can be translated as conditional probabilities giving inequalities like
\begin{equation}
\label{SingIneq1Cond}
\begin{split}
P_{\text{nc}}(D_1^A=1| D_0^B=1) - P_{\text{nc}}(T_0^A=a_0|D_0^B=1)& \\- P_{\text{nc}}(T_1^A=a_1|D_0^B=1)& \leq 0,
\end{split}
\end{equation}
which gives larger violations, since QM predicts $P(D_1^A=1|
D_0^B=1) = \frac{1}{9}$, allowing for $\epsilon' <\frac{1}{27}
\approx 0.037$. Another interesting point is that inequality
\eqref{SingIneq1Cond} only involves one question and one answer
in part $B$: $D_0^B=1$, and an experiment would only involve
three possible measurements in part $A$, with coincidence
countings of each of these three with one and the same
condition in $B$. It is also stressed in the supplementary
material \cite{suppmat} how much this situation resembles the
first example discussed in this Letter.

In this Letter we have established experimentally testable
contextual inequalities, based on the building block of
Kochen-Specker demonstration of the impossibility of
assigning noncontextual answers to all possible predictions
of quantum mechanics. We hope that this proposal will definitely close the
debate on the physical relevance and experimental testability
of the Kochen-Specker theorem, and induce a new generation of
loophole-free experiments.


\begin{acknowledgments}
This work was stimulated by a discussion with A. Zeilinger. The
final work benefited from a discussion with D. Cavalcanti. A.C.
acknowledges support from Capes and ICEx (Belo Horizonte,
Brazil), the Spanish MICINN Project No.\ FIS2008-05596, and the
Wenner-Gren Foundation. M.T.C. acknowledges support from
Brazilian CNPq, Fapemig, and INCT-IQ.
\end{acknowledgments}



\section{Supplementary material}


We start by proving the first Lemma stated in the Letter.


{\em Lemma 1.---}Consider $D_j$ as dichotomic observables
{(with possible outcomes $1$ and $0$)}, $T_j$ as trichotomic
observables {(with possible outcomes $a_j$, $b_j$, and $c_j$)},
with $j=0,1$ and labels $A$ and $B$ corresponding to the
respective party. Then, for any realistic theory, the following
inequality holds:
\begin{subequations}\label{BHinequality}
\begin{equation}
{\cal K} \le 0,
\end{equation}
where
\begin{equation}
\begin{split}
{\cal K}=&\sum_{j \neq k=0}^{1} P(D_j^A=1,D_k^B=1) -
 P(D_0^A=1,T_k^B=a_k)\\
 &-P(T_j^A=a_j,D_0^B=1)-P(D_1^A=1,T_k^B=b_k)\\
 &-P(T_j^A=b_j,D_1^B=1)-P(T_j^A=a_j,T_k^B=b_k)\\
 &-P(T_j^A=a_j,T_j^B=b_j)-P(T_j^A=b_j,T_j^B=a_j)\\
 &-P(T_0^A=a_0,T_0^B=c_0)-P(T_0^A=c_0,T_0^B=a_0)\\
 &-P(T_0^A=b_0,T_0^B=c_0)-P(T_0^A=c_0,T_0^B=b_0)\\
 &-P(T_j^A=c_j,T_k^B=c_k),
\end{split}
\end{equation}
\end{subequations}
where, $P(D_0^A=1,T_0^B=a_0)$ denotes the joint probability of
obtaining the results $1$ and $a_0$ for $D_0^A$ and $T_0^B$,
respectively.


{\em Proof.---}Each probability in \eqref{BHinequality} can be
written as a sum of a set of probabilities for complete hidden
variable states. In our scenario there are four dichotomic
observables, $D_0^A$, $D_1^A$, $D_0^B$, and $D_1^B$, and four
trichotomic observables, $T_0^A$, $T_1^A$, $T_0^B$, and
$T_1^B$. The proof is a bookkeeping exercise showing that each
term that adds to the positive term $P(D_j^A=1,D_k^B=1)$ also
appears (with the right multiplicity) at some of the negative
terms, implying ${\cal K} \le 0$. We will need some notation. A
complete hidden variable state will be given by
\begin{equation}
\label{HVstate}
\left(d_0^A,d_1^A,t_0^A,t_1^A,d_0^B,d_1^B,t_0^B,t_1^B
\right) .
\end{equation}
Any summation without defined index denotes summation over all
the unspecified symbols. With this convention, the positive
contributions for $\cal{K}$ are given by
\begin{equation}
\label{p11}
\begin{split}
\sum_{j \neq k=0}^{1} P(D_j^A=1,&D_k^B=1) \\
 =& \sum P(
 1,d_1^A,t_0^A,t_1^A,d_0^B,1,t_0^B,t_1^B) \\
 & + \sum P(
 d_0^A,1,t_0^A,t_1^A,1,d_1^B,t_0^B,t_1^B).
 \end{split}
\end{equation}
namely, $2\times 2^23^4 = 648$ terms, where $3^4=81$ appear
with multiplicity two. Let us begin by chasing all terms $(
 1,d_1^A,t_0^A,t_1^A,d_0^B,1,t_0^B,t_1^B)$.
The first negative term will cancel all terms like $(
1,d_1^A,t_0^A,t_1^A,d_0^B,1,a_0,t_1^B)$ and $(
1,d_1^A,t_0^A,t_1^A,d_0^B,1,t_0^B,a_1)$, so one eliminates all
terms with $t_0^B = a_0$ or $t_1^B = a_1$. It remains to be
canceled all terms such that the pair $(t_0^B,t_1^B)$ be given
by $(b_0,b_1)$, $(b_0,c_1)$, $(c_0,b_1)$, and $(c_0,c_1)$. The
term $P(T_j^A=b_j,D_1^B=1)$ cancels terms where $t_0^A=b_0$ or
$t_1^A=b_1$. But we still have to look for the $64$ terms
$(1,d_1^A,a_0/c_0,a_1/c_1,d_0^B,1,b_0/c_0,b_1/c_1)$. The term
$P(T_j^A=a_j,T_k^B=b_k)$, with $j=0$, cancels all remaining
terms with $t_0^A=a_0$, except those with $t_1^B=c_1$. Some of
these terms are included in $P(T_0^A=a_0,T_0^B=c_0)$, and we
are now concerned only with the terms
$(1,d_1^A,a_0,a_1/c_1,d_0^B,1,b_0,c_1)$ and
$(1,d_1^A,c_0,a_1/c_1,d_0^B,1,b_0/c_0,b_1/c_1)$. On the same
way, but with $j=1$, $t_1^A=a_1$ with $t_0^B=b_0$ goes away,
leaving $(1,d_1^A,a_0,c_1,d_0^B,1,b_0,c_1)$,
$(1,d_1^A,c_0,a_1,d_0^B,1,c_0,b_1/c_1)$, and
$(1,d_1^A,c_0,c_1,d_0^B,1,b_0,b_1/c_1)$. The first of these
terms goes with $P(T_j^A=a_j,T_j^B=b_j)$, the last goes with
$P(T_0^A=c_0,T_0^B=b_0)$, while the remaining reduces to
$(1,d_1^A,c_0,a_1,d_0^B,1,c_0,b_1)$, due to
$P(T_j^A=c_j,T_k^B=c_k)$, which is finally canceled by
$P(T_j^A=a_j,T_j^B=b_j)$.

Now we can follow a similar strategy for the terms
$(d_0^A,1,t_0^A,t_1^A,1,d_1^B,t_0^B,t_1^B)$. Starting from
$P(T_j^A=a_j,D_0^B=1)$, which eliminates all terms with
$t_0^A=a_0$ or $t_1^A=a_1$, and $P(D_i^A=1,T_k^B=b_k)$, which
eliminates $t_0^B=b_0$ and $t_1^B=b_1$, we are left with
$(d_0^A,1,b_0/c_0,b_1/c_1,1,d_1^B,a_0/c_0,a_1/c_1)$. The term
$P(T_j^A=b_j,T_j^B=a_j)$, with $j=0$, reduces the task to
$(d_0^A,1,b_0,b_1/c_1,1,d_1^B,c_0,a_1/c_1)$ and
$(d_0^A,1,c_0,b_1/c_1,1,d_1^B,a_0/c_0,a_1/c_1)$. The same term,
with $j=1$, reduces even more to
$(d_0^A,1,b_0,b_1,1,d_1^B,c_0,c_1)$,
$(d_0^A,1,b_0,c_1,1,d_1^B,c_0,a_1/c_1)$,
$(d_0^A,1,c_0,b_1,1,d_1^B,a_0/c_0,c_1)$, and
$(d_0^A,1,c_0,c_1,1,d_1^B,a_0/c_0,a_1/c_1)$.
The two first terms are eliminated by $P(T_0^A=b_0,T_0^B=c_0)$,
while $P(T_0^A=c_0,T_0^B=a_0)$ reduces the order two to
$(d_0^A,1,c_0,b_1,1,d_1^B,c_0,c_1)$, and
$(d_0^A,1,c_0,c_1,1,d_1^B,c_0,a_1/c_1)$. The term
$P(T_j^A=c_j,T_k^B=c_k)$ with $j=0$ eliminates the first and
reduces the second to $(d_0^A,1,c_0,c_1,1,d_1^B,c_0,a_1)$,
which is eliminated by the same term with $j=1$.

As already said, $81$ terms appear with twice each in the
positive part. We close the demonstration by proving that they
also appear (at least) twice in the negative contributions.
They are $(1,1,t_0^A,t_1^A,1,1,t_0^B,t_1^B)$. Let us start by
the terms $(1,1,a_0,t_1^A,1,1,t_0^B,t_1^B)$. They appear once
at $P(T_j^A=a_j,D_0^B=1)$, for $j=0$, and a second time at
$P(D_0^A=1, T_k^B=a_k)$, $k=0$, if $t_0^B=a_0$, at
$P(T_j^A=a_j,T_j^B=b_j)$, $j=0$, if $t_0^B=b_0$, and at
$P(T_0^A=a_0,T_0^B=c_0)$, in case $t_0^B-c_0$. Analogously, for
the terms $(1,1,b_0,t_1^A,1,1,t_0^B,t_1^B)$, they appear once
at $P(T_j^A=b_j,D_1^B=1)$, and again at
$P(T_j^A=b_j,T_j^B=a_j)$, $j=0$, if $t_0^B=a_0$, at
$P(D_1^A=1,T_k^B=b_k)$, $k=0$, if $t_0^B=b_0$, and at
$P(T_0^A=b_0,T_0^B=c_0)$, if $t_0^B=c_0$. Finally, for the
states $(1,1,c_0,t_1^A,1,1,t_0^A,t_1^A)$, they appear once at
$P(D_0^A=1,T_k^B=a_k)$, $k=1$, if $t_1^B=a_1$, at
$P(D_1^A=1,T_k^B=b_k)$, $k=1$, if $t_1^B=b_1$, and at
$P(T_j^A=c_j,T_k^B=c_k)$, $k=1$, if $t_1^B=c_1$. The second
appearance can be noted at $P(T_0^A=c_0,T_0^B=a_0)$, if
$t_0^B=a_0$, at $P(T_0^A=c_0,T_0^B=b_0)$ if $t_0^B=b_0$, while
for $t_0^B=c_0$ we split again into three terms:
$P(T_j^A,D_0^B)$, $j=1$ for $t_1^A=a_1$,
$P(T_j^A=b_j,D_1^B=1)$, $j=1$, for $t_1^A=b_1$, and
$P(T_j^A=c_j,T_k^B=c_k)$, $j=1$, for $t_1^A=c_1$.
\hfill\endproof


Before proving Lemma 2, let us list all noncontextual
hidden-variable states giving answers to the four questions
which one can ask for one party, subjected to the restrictions:
\begin{subequations}
 \label{compat}
\begin{align}
 &T_j=a_j \Rightarrow D_0=0, \\
 &T_j=b_j \Rightarrow D_1=0, \\
 &T_j=c_j \Rightarrow T_k \neq c_k\;(j\neq k).
\end{align}
\end{subequations}
As stated, only the following $14$ out of the $36$ realistic
states are noncontextual:
\begin{equation}
\begin{split}
 &(0,q_1,a_0,a_1), (0,0,a_0,b_1), (0,q_1,a_0,c_1),(0,0,b_0,a_1),\\
 &(q_0,0,b_0,b_1), (q_0,0,b_0,c_1), (0,q_1,c_0,a_1), (q_0,0,c_0,b_1).
\end{split}
\end{equation}
When considering two parties, the cross product must be
applied, resulting in $14^2 = 196$ noncontextual states.

We are now ready to state and prove a generalized version of
Lemma 2 in the text.


{\em Lemma 2'.---}Consider {$D_j$ and $T_j$ as in Lemma 1.}
Moreover, consider these questions to obey, for each party,
conditions \eqref{compat}. The following inequalities hold:
\begin{subequations}
\label{SingIneq1}
\begin{align}
P_{\text{nc}}(D_0^A=1, D_1^B=1) &- P_{\text{nc}}(D_0^A=1, T_0^B=a_0) \nonumber \\ &- P_{\text{nc}}(D_0^A=1, T_1^B=a_1) \leq 0,\label{SingIneq1a}\\
P_{\text{nc}}(D_1^A=1, D_0^B=1) &- P_{\text{nc}}(T_0^A=a_0,D_0^B=1) \nonumber \\ &- P_{\text{nc}}(T_1^A=a_1,D_0^B=1) \leq 0,\\
P_{\text{nc}}(D_0^A=1, D_1^B=1) &- P_{\text{nc}}(D_0^A=1, T_0^B=b_0) \nonumber \\ &- P_{\text{nc}}(D_0^A=1, T_1^B=b_1) \leq 0,\\
P_{\text{nc}}(D_1^A=1, D_0^B=1) &- P_{\text{nc}}(T_0^A=b_0,D_0^B=1) \nonumber \\ &- P_{\text{nc}}(T_1^A=b_1,D_0^B=1) \leq 0.
\end{align}
\end{subequations}
Where $P_{\text{nc}}$ is used to remind that only noncontextual
hidden-variable states are taken into account.


{\em Proof.---}Let us prove inequality \eqref{SingIneq1a} in
detail. The others are analogous. Note that only three
one-party states allow for $D_0^A=1$: $(1,0,b_0,b_1)$,
$(1,0,b_0,c_1)$, and $(1,0,c_0,b_1)$. In the same way, only
three states of party $B$ will allow for $D_1^B=1$:
$(0,1,a_0,a_1)$, $(0,1,a_0,c_1)$, and $(0,1,c_0,a_1)$. So we
have $9$ noncontextual states contributing for
$P_{\text{nc}}(D_0^A=1,D_1^B=1)$. But note that six of these
states have $t_0^B=a_0$, and the other three have $t_1^B=a_1$,
so once again all terms which add for the positive part also
appear in the negative one. As probabilities are non-negative,
the resulting inequality applies. \hfill\endproof


To see that inequality \eqref{SingIneq1a} is tight, just use
state $(1,0,b_0,b_1,0,1,a_0,c_1)$, while
$(1,0,b_0,b_1,0,1,a_0,a_1)$ is a realistic contextual state (in
the sense that it is an assignment of values which does not
obey the assumption of noncontextuality) which violates it.

It is now time to prove the following.


{\em Lemma 3.---}With the same conditions of Lemma 2,
\begin{equation}
\label{SingIneq2}
\begin{split}
P_{\text{nc}}(D_0^A=1, D_1^B=1) &-\sum_j P_{\text{nc}}(T_j^A=b_j, T_j^B=a_j) \\
 &- \sum _{j\neq k} P_{\text{nc}}(T_j^A=c_j, T_k^B=c_k) \leq 0.
\end{split}
\end{equation}


{\em Proof.---}It is again necessary to cancel out the same
nine terms which contribute to
$P_{\text{nc}}(D_0^A=1,D_1^B=1)$. Four of them have $t_0^A=b_0$
and $t_0^B=a_0$. Also four of them have $t_1^A=b_1$ and
$t_1^B=a_1$, but one was taken twice. The remaining two terms
are canceled by the second summation term, since one has
$(T_0^A=c_0, T_1^B=c_1)$ and the other $(T_1^A=c_1,T_0^B=c_0)$.
\hfill\endproof

Again, \eqref{SingIneq2} is tight with respect to noncontextual
states, as the state $(1,0,b_0,c_1,0,1,a_0,c_1)$ shows, and can
be violated with realistic contextual ones like
$(1,0,a_0,a_1,0,1,b_0,b_1)$.


{\em Quantum violation.---}Now let us discuss in more details
how the quantum violation of the presented inequalities is
directly related to the first example discussed in the Letter.
For this, we will focus on the inequality
\begin{equation}
\label{SingIneq1Cond}
\begin{split}
P_{\text{nc}}(D_1^A=1| D_0^B=1) &- P_{\text{nc}}(T_0^A=a_0|D_0^B=1) \\
&- P_{\text{nc}}(T_1^A=a_1|D_0^B=1) \leq 0.
\end{split}
\end{equation}
Since every term is conditioned to $D_0^B=1$, we can consider a
projective measurement on part $B$ with the result
corresponding to $|{f}\rangle\langle{f}|$ appearing. Since the
conditional state of
\begin{equation}
|\psi\rangle = \frac{1}{\sqrt{3}} \left(|02\rangle - |11\rangle + |20\rangle\right)
\label{singlete}
\end{equation}
given that $|f\rangle _B$ was obtained is $|i\rangle _A$, one
is faced with the quantum probabilities
\begin{subequations}
\label{QMcondProbs}
\begin{align}
&P(D_1^A=1|D_0^B=1)=|\langle f|i\rangle |^2 = \frac{1}{9},\\
&P(T_0^A=a_0|D_0^B=1)=|\langle a_0|i\rangle |^2 =0,\\
&P(T_1^A=a_1|D_0^B=1)=|\langle a_1|i\rangle |^2 =0,
\end{align}
\end{subequations}
from where not only the violation is evident, but also the
relation with the previous discussed example must be clear.


\end{document}